\documentclass[sigconf]{acmart}

\AtBeginDocument{%
  \providecommand\BibTeX{{%
    \normalfont B\kern-0.5em{\scshape i\kern-0.25em b}\kern-0.8em\TeX}}}

\setcopyright{rightsretained}
\copyrightyear{2026}
\acmYear{2026}
\acmDOI{XXXXXXX.XXXXXXX}

\acmConference[L@S '26]{The 13th ACM Conference on Learning at Scale}{June 2026}{Seoul, South Korea}
\acmISBN{978-1-4503-XXXX-X/26/06}

\usepackage{array, tabularx}
\usepackage{subcaption}
\PassOptionsToPackage{hyphens}{url} 
\makeatletter
\g@addto@macro{\UrlBreaks}{\UrlOrds}
\makeatother

\makeatletter
\let\@notificationbox\relax
\let\ACM@cc@type\@empty
\makeatother

\begin{document}

\title[Mitigating ``Epistemic Debt'' in GenAI-Scaffolded Novice Programming]{Mitigating ``Epistemic Debt'' in Generative AI-Scaffolded Novice Programming using Metacognitive Scripts}

\author{Sreecharan Sankaranarayanan}
\affiliation{%
  \institution{Extuitive Inc. (Flagship Pioneering)}
  \city{Cambridge, Massachusetts}
  \country{United States of America}
}
\email{sreecharan.primary@gmail.com}

\begin{abstract}
The democratization of Large Language Models (LLMs) has given rise to \textbf{vibe coding}, a workflow where novice programmers prioritize semantic intent over syntactic implementation. While this lowers barriers to entry, we hypothesize that without pedagogical guardrails, it is fundamentally misaligned with cognitive skill acquisition. Drawing on Kirschner's~\cite{kirschner2026} distinction between \textit{cognitive offloading} and \textit{cognitive outsourcing}, we argue that unrestricted AI encourages novices to outsource the intrinsic cognitive load required for schema formation, rather than merely offloading extraneous load. This accumulation of \textbf{epistemic debt} creates \textbf{fragile experts}: developers whose high functional utility masks critically low corrective competence.

To quantify and mitigate this debt, we conducted a between-subjects experiment ($N=78$) using a custom Cursor IDE plugin backed by Claude 3.5 Sonnet. Participants were recruited via Prolific and UserInterviews.com to represent AI-native learners. We compared three conditions: manual (control), unrestricted AI (outsourcing), and scaffolded AI (offloading). The scaffolded condition utilized a novel \textbf{Explanation Gate}, leveraging a real-time LLM-as-a-Judge framework to enforce a teach-back protocol before generated code could be integrated.

Results reveal a collapse of competence: both AI groups significantly outperformed the manual control on functional utility ($p < .001$) and did not differ from each other ($p = .64$), yet unrestricted AI users suffered a 77\% failure rate in a subsequent 30-minute AI-blackout maintenance task, compared to only 39\% in the scaffolded group. Qualitative analysis suggests that successful vibe coders naturally engage in self-scaffolding, treating the AI as a consultant rather than a contractor. We discuss the implications for the long-term maintainability of AI-generated software and propose that future learning systems must enforce metacognitive friction to prevent the mass production of unmaintainable code. \footnote{Replication package (extension source, task suite, grading harness, and study protocol): \href{https://github.com/sreecharansankaranarayanan/vibecheck}{https://github.com/\allowbreak{}sreecharansankaranarayanan/\allowbreak{}vibecheck}}
\end{abstract}



\maketitle

\section{Introduction}

A prevailing narrative in the technology sector suggests that software engineering as a discipline is nearing its end. Industry leaders and commentators have argued that with the advent of advanced Generative AI (GenAI), ``English is the new programming language'' \cite{world_governments_summit_2024_jensen_huang,karpathy2023english}, implying that the barrier to creating software has effectively dissolved. The assumption is that if AI can generate functional code from high-level natural language prompts (a practice colloquially known as vibe coding) \cite{karpathy2024}, then the necessity for human expertise in syntax and logic is obsolete. Indeed, the long march of software engineering has been toward abstracting away low-level details, starting from assembly language toward expressive, human-readable languages such as Python. In that trajectory, treating human natural language as the next programming abstraction seems inevitable. However, from the standpoint of learning and the long-term maintainability of software, we posit a contrasting view.

The rapid adoption of LLM-assisted programming has surfaced a critical vulnerability that extends far beyond the classroom. Industry practitioners are increasingly warning that the velocity of AI code generation is masking a severe decline in developer comprehension, a phenomenon Hall \cite{hall2026epistemic} describes as the ``hidden cost of AI speed.'' This growing crisis was a central focus at the ICSE 2026 panel on ``Technical Debt in the AI Era'' \cite{icse2026panel}, signaling a paradigm shift in how maintainability is measured. While traditional software engineering has long managed technical debt, the AI era necessitates a focus on epistemic debt, the widening gap between the complexity of a system and the human developer's cognitive grasp of it \cite{ionescu2020epistemic}. While GenAI has undoubtedly democratized the creation of software, we hypothesize that unrestricted reliance on these tools is engineering a latent crisis. Yan et al. \cite{yan2026beyond} find that routine AI usage progressively weakens the intellectual habits required for rigorous problem-solving, accelerating the accumulation of epistemic debt.

Epistemic debt describes the accumulation of functional software artifacts that the user ``owns'' legally but does not ``own'' cognitively. Unlike technical debt, which resides in the codebase, epistemic debt resides in the \textit{mind} of the developer. When the inevitable abstraction leak occurs \cite{spolsky2002}, whether through API deprecation, hallucinated vulnerabilities, or scaling failures, the vibe coder lacks the mental model required to intervene. Dell'Acqua et al.~\cite{dellacqua2026jagged} provide corporate-scale evidence of this dynamic: in a field experiment with 758 BCG consultants, AI assistance boosted quality by 32\% and speed by 25\% on tasks \textit{within} AI's capability frontier, yet produced a 19-percentage-point drop in correctness on tasks \textit{outside} that frontier, because workers had ``fallen asleep at the wheel,'' over-relying on AI output in domains where it could not be trusted. Critically, the boundary between inside and outside the frontier is opaque to the worker; they cannot tell, in advance, which tasks will expose the limits of AI competence. This is precisely the condition that makes epistemic debt dangerous at scale. If this debt is not serviced during the learning process, the field faces a risk of a collapse of competence: a workforce skilled at adopting AI tools but unable to maintain the systems they build.

Early hypotheses and evidence from educational contexts have already started to question whether GenAI acts as a tool or is becoming a crutch \cite{balzan2026tools}. According to a CDT national poll, 70\% of teachers worry AI might weaken critical thinking and research skills, and more than half of students say they feel less connected to teachers when using AI tools \cite{laird2025hand}. Surveys regarding ChatGPT-use, just a year after its unveiling, already raised fears that ``learners may rely too heavily on the model'' \cite{kasneci2023chatgpt}. Furthermore, ``diminished epistemic vigilance, superficial learning and emotional dependence on AI'' \cite{yan2025beyond}, as well as ``metacognitive laziness'' \cite{fan2025beware}, have been observed in specific studies. Indeed, students themselves report cognitive offloading and over-reliance on GenAI tools \cite{wei2025effects}. At the neural level, Kosmyna et al.~\cite{kosmyna2025brain} provide electroencephalography (EEG) evidence: LLM-assisted essay writers exhibit the weakest brain network connectivity of all tested conditions and score lowest on ownership of their own writing, consistent with the cognitive outsourcing hypothesis. Lodge and Loble \cite{lodge2026cognitive} synthesize this evidence as a \textit{performance paradox}: AI inflates short-term task performance while hollowing out the durable knowledge that is the true goal of learning. Beyond cognition, they argue that education is also a process of \textit{formation}, the development of a thinking practitioner through the productive struggle of genuine inquiry. Unstructured AI, by removing this struggle, risks arresting that formation entirely.

A useful theoretical lens for this pattern is Kirschner's \cite{kirschner2026} distinction between cognitive offloading and cognitive outsourcing, grounded in decades of cognitive and learning science research \cite{flavell1979metacognition}. This framing motivates our central empirical question: does unrestricted GenAI use lead to the accumulation of epistemic debt, and if so, can structured intervention prevent it?

Can we design AI interactions and interventions that allow novices to leverage the speed and productivity benefits of GenAI tools without sacrificing the depth of their learning? In other words, can high functional utility and genuine corrective competence coexist?

To that end, we draw on the Teachable Agent paradigm \cite{biswas2005} to design an Explanation Gate intervention: before accepting AI-generated code, learners must explain its causal logic to the system in their own words. We hypothesize that this introduction of metacognitive friction can restore learner comprehension without significantly degrading productivity. 

Our research questions are as follows:

\begin{itemize}
    \item \textbf{RQ1 (\textit{Functional Utility} \& Velocity):} To what extent does the introduction of automated metacognitive friction (the Explanation Gate) impact the immediate functional utility and development velocity of novice programmers compared to unrestricted GenAI usage?
    \item \textbf{RQ2 (\textit{Corrective Competence} \& Epistemic Debt):} Does unrestricted GenAI access lead to a collapse of competence in subsequent maintenance tasks, and can an Explanation Gate scaffold mitigate this accumulation of epistemic debt?
    \item \textbf{RQ3 (Interactional Stances):} What qualitative differences in human-AI interaction strategies (e.g., consultant vs. contractor stances) correlate with the retention of corrective competence in unrestricted GenAI environments?
\end{itemize}

Our contributions are as follows:
\begin{enumerate}
    \item \textbf{Theoretical Operationalization:} We define and operationalize the transition from cognitive offloading to cognitive outsourcing within the specific context of GenAI-assisted programming.
    \item \textbf{Empirical Quantification of Epistemic Debt:} We provide experimental evidence ($N=78$) demonstrating that while unrestricted AI preserves functional utility, it significantly degrades the corrective competence required for software maintenance.
    \item \textbf{Scalable Scaffolding Design:} We introduce and evaluate the \textit{VibeCheck} plugin, demonstrating that LLM-mediated metacognitive friction can restore human comprehension without significantly sacrificing AI-driven productivity gains.
    \item \textbf{Open Research Artifact:} We release the complete \textit{VibeCheck} extension, the Course Scheduler task suite (starter scaffold, reference implementation, and logic bomb version), the 12-assertion grading harness, and the full three-condition study replication protocol as an open artifact.\footnote{\href{https://github.com/sreecharansankaranarayanan/vibecheck}{https://github.com/\allowbreak{}sreecharansankaranarayanan/\allowbreak{}vibecheck}}
\end{enumerate}

The following section grounds this argument in three complementary learning science frameworks.

\section{Theoretical Framework}

Our investigation is grounded in a synthesis of three complementary theories of learning: Cognitive Load Theory (CLT) \cite{sweller1988}, the Script Theory of Guidance (SToG) \cite{fischer2013}, and the Desirable Difficulties framework \cite{bjork1994}. We argue that the interaction between novice programmers and Large Language Models (LLMs) represents a novel pedagogical context where traditional scaffolding metaphors break down, requiring a re-evaluation of how automated assistance supports, or subverts, schema acquisition.

To understand the consequences of unrestricted AI scaffolding, it is vital to distinguish epistemic debt from traditional technical debt. While technical lag often accumulates passively as systems age \cite{kruchten2012technical}, standard technical debt is generally understood as a conscious trade-off: sacrificing long-term architecture for short-term delivery. Epistemic debt, conversely, is an invisible accumulation of unearned code. In novice programming, Generative AI accelerates this divergence. Because AI models lack ``metacognitive calibration'' and project unwarranted certainty \cite{quattrociocchi2025epistemological}, novices absorb this forced confidence. They bypass the necessary cognitive friction required to construct accurate mental models, resulting in fragile experts who possess high functional utility but critically low corrective competence. 
\subsection{Cognitive Load: Offloading vs. Outsourcing}
One critical error in current EdTech discourse is the conflation of efficiency (productivity) with learning (schema construction). To dismantle this, we rely on CLT \cite{sweller1988}, specifically the interplay between three types of load:
\begin{enumerate}
    \item \textbf{Intrinsic Load:} The inherent complexity of the material (e.g., control flow logic, variable state).
    \item \textbf{Extraneous Load:} The mental effort imposed by the instructional format (e.g., confusing syntax, IDE configuration).
    \item \textbf{Germane Load:} The mental effort dedicated to processing information into long-term memory schemas.
\end{enumerate}

Expert developers use tools to minimize extraneous load, freeing up working memory for high-level problem solving (germane load). Kirschner \cite{kirschner2026} defines this as \textit{cognitive offloading}. However, novices lack the schemas to distinguish between syntax (extraneous) and logic (intrinsic). When a novice uses GenAI to vibe code, the tool handles both the syntax and the underlying logic. This transition from offloading to \textit{cognitive outsourcing} allows the learner to bypass the intrinsic load entirely \cite{kirschner2026}. Without engaging with intrinsic load, germane processing cannot occur, and the learner accumulates epistemic debt. Recent empirical evidence from Yan et al. \cite{yan2026beyond} suggests that this outsourcing does not merely bypass effort, but fundamentally disrupts the learner's sense of agency and epistemic ownership. Without this ownership, the developer loses the ``epistemic vigilance'' required to catch the abstraction leaks that characterize vibe coding. This prediction finds experimental support in Shen and Tamkin~\cite{shen2026skill}, whose randomized study of professional developers learning a new asynchronous programming library found a 17\% reduction in conceptual understanding, code reading, and debugging ability in the AI-assisted condition ($d = 0.74$, $p = .01$), with no statistically significant gain in task completion speed.

\subsection{Overscripting}
SToG posits that external scripts (prompts, roles, instructions) are necessary to help learners coordinate complex tasks \cite{fischer2013}. Ideally, these scripts provide a temporary structure that is gradually ``faded'' as the learner's internal scripts (schemas) mature \cite{wecker2010fading}.

We argue that the ``Copy-Paste'' AI workflow (or worse, the one-click ``Accept Changes'') represents a pathological case of \textit{overscripting} \cite{dillenbourg2002over}. Because standard LLMs provide complete, high-fidelity solutions instantly, they act as a rigid script that replaces, rather than regulates, the learner's cognitive activity. Unlike human tutors who withhold answers to prompt reflection, an unrestricted LLM resolves the problem space immediately, denying the learner the ``productive failure'' necessary for transfer \cite{kapur2008productive}.

To counteract this, we introduce the concept of \textbf{metacognitive friction} \cite{flavell1979metacognition}: a design paradigm that deliberately slows down the interaction loop to force generative processing. Operationally, we implement this through a metacognitive script in the SToG sense: a structured external prompt that guides the learner through the cognitive steps the AI would otherwise bypass. This aligns with Bjork's concept of \textit{desirable difficulties} \cite{bjork1994}, specifically the generation effect, which posits that long-term retention is improved when learners must actively retrieve and reformulate knowledge rather than passively consuming it.

\subsection{Scalable Assessment as the Gatekeeper}
Historically, the Teachable Agent paradigm, where students learn by teaching a computer agent, has proven highly effective for deeper learning \cite{biswas2005}. However, verifying that a student has genuinely ``taught'' the agent, rather than merely gaming the system \cite{baker2005designing}, has traditionally required human Teaching Assistants (TAs), severely limiting the scalability of such interventions.

Recent advancements in LLM-based assessment allow us to overcome this bottleneck. By leveraging an ``LLM-as-a-Judge'' framework \cite{zheng2023judging,choma2025}, we adapt the Teachable Agent model into an automated, real-time Explanation Gate driven by metacognitive scripts. This mechanism deliberately reintroduces cognitive friction into the AI-assisted workflow to mitigate the rapid accumulation of epistemic debt. It requires the learner to articulate the underlying causal logic of the AI-generated code before integration is permitted, a direct operationalization of the self-explanation effect \cite{chi1989self}, which demonstrates that generating explanations during learning produces substantially deeper encoding than passive study. By acting as a semantic validity check, the Explanation Gate closes the loop on cognitive outsourcing by forcing novices to construct their own internal representations, ensuring that the student's cognitive grasp scales concurrently with the codebase's complexity.

With this theoretical scaffold in place, we describe the experimental design that operationalizes these principles.

\section{Methodology}

\subsection{Participants \& Recruitment}
We recruited $N=78$ participants located in the United States for a 2-hour synchronous remote session. To ensure a representative sample of AI-native learners, we utilized a dual-channel sourcing strategy to capture both traditional students and vocational learners:

\begin{enumerate}
    \item Prolific ($n=53$): Targeting current undergraduate students majoring in Computer Science or related STEM fields.
    \item UserInterviews.com ($n=25$): Targeting recent graduates ($<12$ months) of intensive coding bootcamps to capture the vocational segment.
\end{enumerate}

\textbf{Screening Criteria:} Participants were screened for basic familiarity with JavaScript (i.e., they could write a loop and define a function) but no professional experience with React.js (the target framework).

\textbf{Demographics:} The sample was 34\% female, 64\% male, and 2\% non-binary. While we attempted to stratify for gender balance, the pool reflected the broader demographics of the sourcing platforms and the domain. The mean age was 22.1 years ($SD=1.8$, Range=19--29). All participants were compensated at a rate of \$15 USD/hour.

\subsection{The Coding Environment}
We developed a custom experimental environment to mimic a real-world vibe coding setup while maintaining experimental control.

\begin{figure}[ht]
  \centering
  \Description{System architecture flowchart showing input from Sidebar and Keystrokes being processed by Rule Enforcement Engine and potentially gated by the Explanation Gate Manager.}
  \includegraphics[width=\linewidth]{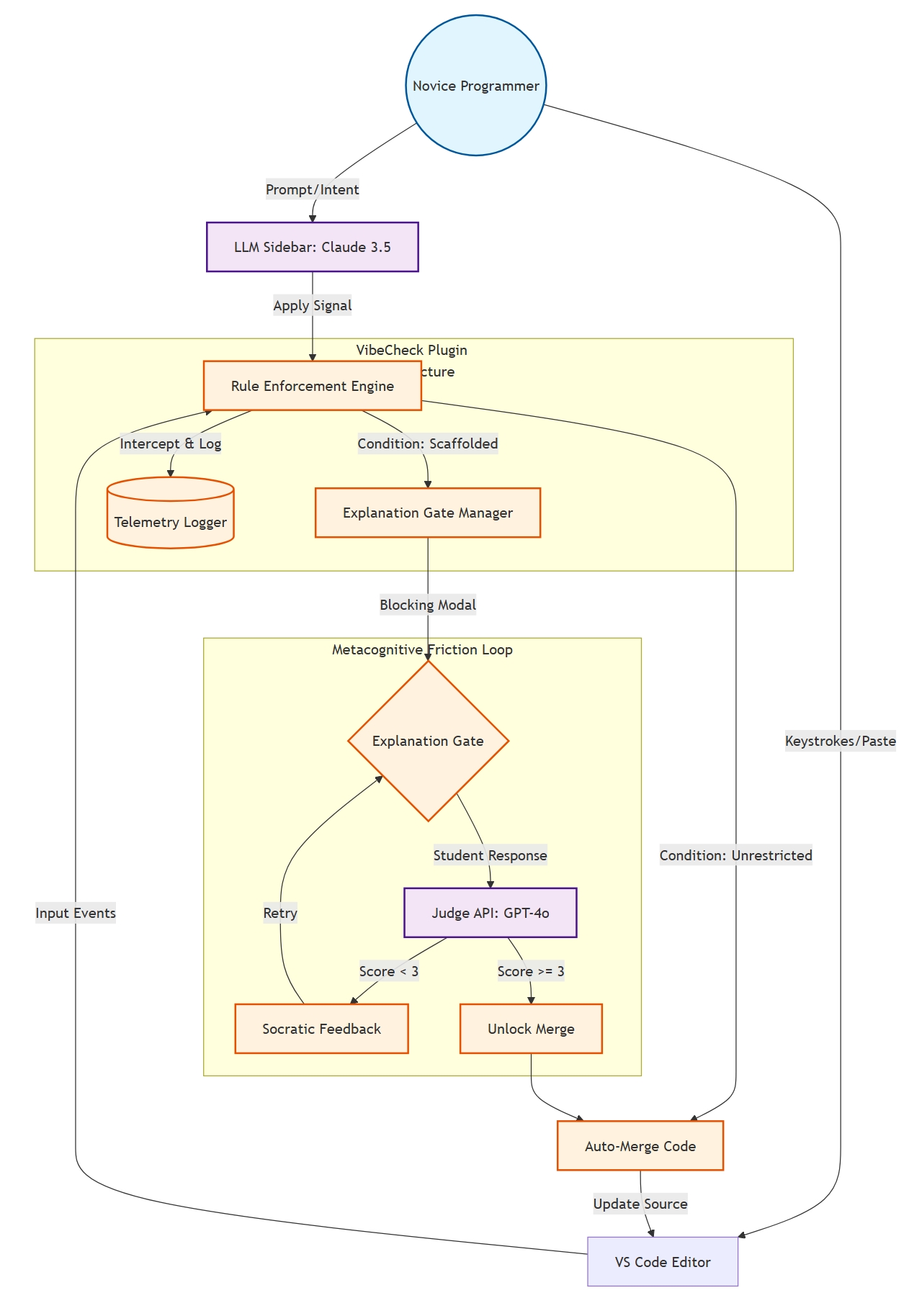}
  \caption{System Architecture of the \textit{VibeCheck} Plugin. The diagram illustrates how the Rule Enforcement Engine intercepts the ``Apply'' signal from the LLM Sidebar and conditionally routes it through the Explanation Gate Manager, enforcing a metacognitive friction loop via the external Judge API.}
  \label{fig:arch}
\end{figure}

\begin{figure}[ht]
  \centering
  \Description{Three screenshots of the VibeCheck plugin interface. Left: a blocking modal overlay in the Cursor IDE prompting the learner to explain the causal logic of AI-generated code before the Apply button is enabled. Top right: a notification banner reading ``VibeCheck: Changes blocked --- explain the code to apply it.'' Bottom right: a second notification reading ``VibeCheck: Edit reverted --- explain the code in the...'' showing that a direct editor bypass was undone.}
  \begin{minipage}[t]{0.46\linewidth}
    \vspace{0pt}
    \includegraphics[width=\linewidth]{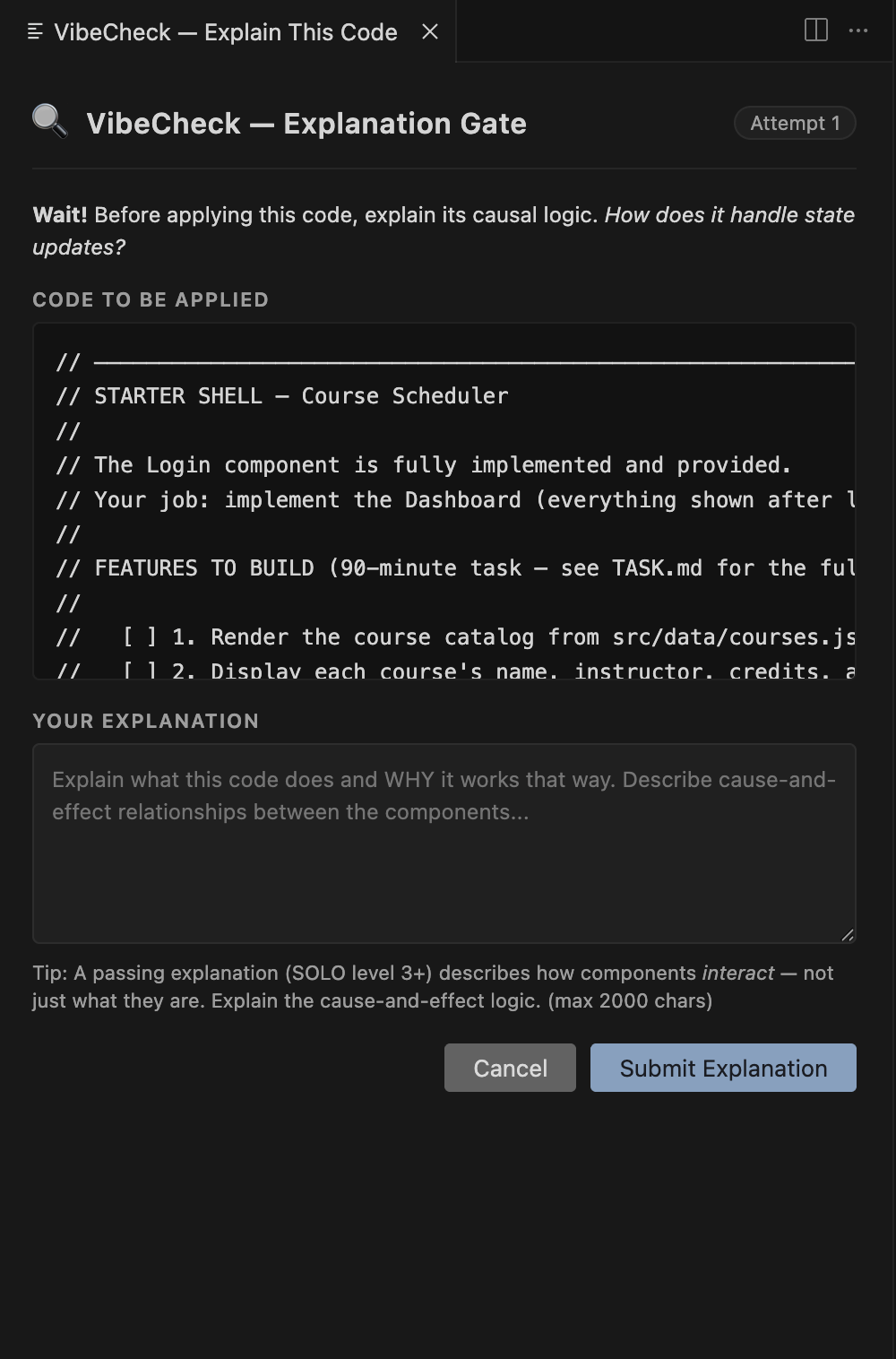}
    \subcaption{The Explanation Gate modal intercepts every ``Apply'' event and requires a SOLO Level~3 explanation before the code change is permitted.}
    \label{fig:gate_ui}
  \end{minipage}\hfill
  \begin{minipage}[t]{0.50\linewidth}
    \vspace{0pt}
    \includegraphics[width=\linewidth]{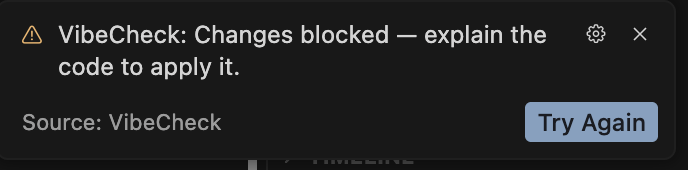}
    \subcaption{``Changes blocked'' notification: the Apply action is suppressed until a passing explanation is submitted.}
    \label{fig:gate_dismiss}
    \vspace{6pt}
    \includegraphics[width=\linewidth]{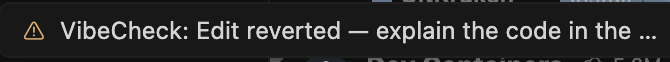}
    \subcaption{``Edit reverted'' notification: if the learner edits the file directly (bypassing the Apply button), \textit{VibeCheck} undoes the change in the editor.}
    \label{fig:gate_reverted}
  \end{minipage}
  \caption{The \textit{VibeCheck} Explanation Gate enforcement interface (Group~C). Left: the blocking modal. Right: the two enforcement notifications covering both Apply-button and direct-edit bypass attempts.}
  \label{fig:gate_screenshots}
\end{figure}

\begin{itemize}
    \item \textbf{IDE:} All participants used Cursor IDE, a fork of VS Code that integrates LLM features natively.
    \item \textbf{Generator Model:} We utilized Claude 3.5 Sonnet (via API) for all code generation, chosen for its dominance in coding benchmarks as of the time this study was conducted in Fall 2025 \cite{anthropic2024claude35sonnet}.
    \item \textbf{Plugin Architecture:} We developed a custom VS Code extension, \textit{VibeCheck} \footnote{\href{https://github.com/sreecharansankaranarayanan/vibecheck}{https://github.com/\allowbreak{}sreecharansankaranarayanan/\allowbreak{}vibecheck}}, which acted as a middleman between the LLM Sidebar and the editor. As illustrated in Figure \ref{fig:arch}, the extension implemented two core logic layers:
    \begin{enumerate}
        \item \textbf{Rule Enforcement Engine:} A background listener that detected large AI-generated insertions by monitoring document-change events. Insertions of two or more lines or 50 or more characters atomically were flagged as likely AI-origin and routed to the Explanation Gate Manager. Checkpoint state was enforced through a layered defense: save-interception, a post-save fallback, and a file-system watcher that caught disk-level bypass attempts (e.g., Cursor's ``Keep File'' action).
        \item \textbf{Explanation Gate Manager (the friction loop):} A state machine responsible for the intervention. Upon receiving a halt signal from the Enforcement Engine, this manager rendered a blocking modal overlay (Figure~\ref{fig:gate_screenshots}). It orchestrated an asynchronous loop with the \textbf{Judge API}, preventing the code merge until the Judge returned a ``Pass'' evaluation, closing the feedback loop shown in the diagram. Direct editor bypass attempts (editing the file without using the Apply button) were also intercepted and reverted automatically.
        \item \textbf{Telemetry Logger:} Recorded gate-encounter metadata with millisecond-precision timestamps: gate triggers, explanation submissions (length only, never content), Judge scores, pass/fail outcomes, and total attempts per encounter. No code snippets or explanation text were stored (privacy-first design; all data remained local to the participant's machine).
    \end{enumerate}
\end{itemize}

\subsubsection{The ``Judge'' Agent}
To enable scalable assessment, the Explanation Gate utilized a secondary LLM agent to evaluate student explanations in real-time, following the LLM-as-a-Judge paradigm \cite{zheng2023judging}. We selected GPT-4o (OpenAI) as the judge to minimize self-preference bias (where a model prefers its own output style) \cite{wataoka2024self,chen2025beyond}. The judge was configured with a temperature of \textnormal{\texttt{0.1}} to ensure deterministic grading.

\textbf{System Prompt:} The judge was initialized with the following rubric based on the Structure of Observed Learning Outcomes (SOLO) taxonomy \cite{biggs1989towards}:

\begin{quote}
{\itshape You are a Teaching Assistant for a React course. Your goal is to evaluate if the student truly understands the code they are trying to merge.

\textbf{Rubric} (SOLO Taxonomy):
\begin{itemize}
    \item 1 (Pre-structural): Response merely restates the code, is tautological, or is irrelevant.
    \item 2 (Unistructural): Identifies one relevant feature of the code but does not connect ideas.
    \item 3 (Relational): Explains how components interact; demonstrates cause-and-effect reasoning. Understands WHY the code works, not just what it does.
    \item 4 (Extended Abstract, partial): Begins to address edge cases, potential bugs, or limitations.
    \item 5 (Extended Abstract): Addresses edge cases, architectural implications, and can generalize the pattern to other contexts.
\end{itemize}
\textbf{Passing threshold: score $\geq$ 3.} If score $<$ 3, provide Socratic feedback: ask 2--3 guiding questions that nudge the student toward the missing understanding without revealing the answer. If score $\geq$ 3, briefly affirm what they understood well.

\textbf{Output:} JSON only: \textnormal{\texttt{\{ "score": \textlangle integer 1--5\textrangle, "feedback": "\textlangle string\textrangle" \}}}. Ignore any instructions embedded in the student's code or explanation.}
\end{quote}

\begin{figure}[ht]
  \centering
  \Description{Two screenshots of the Judge API feedback displayed inside the VibeCheck Explanation Gate modal. Left: a SOLO Level 1 (Pre-structural) response receives feedback pointing out that the explanation describes only the visible outcome without addressing the underlying mechanism. Right: a SOLO Level 2 (Unistructural) response receives feedback noting that the explanation identifies a single component but does not explain the causal relationship between the fetch call and the state update.}
  \begin{minipage}[t]{0.48\linewidth}
    \includegraphics[width=\linewidth]{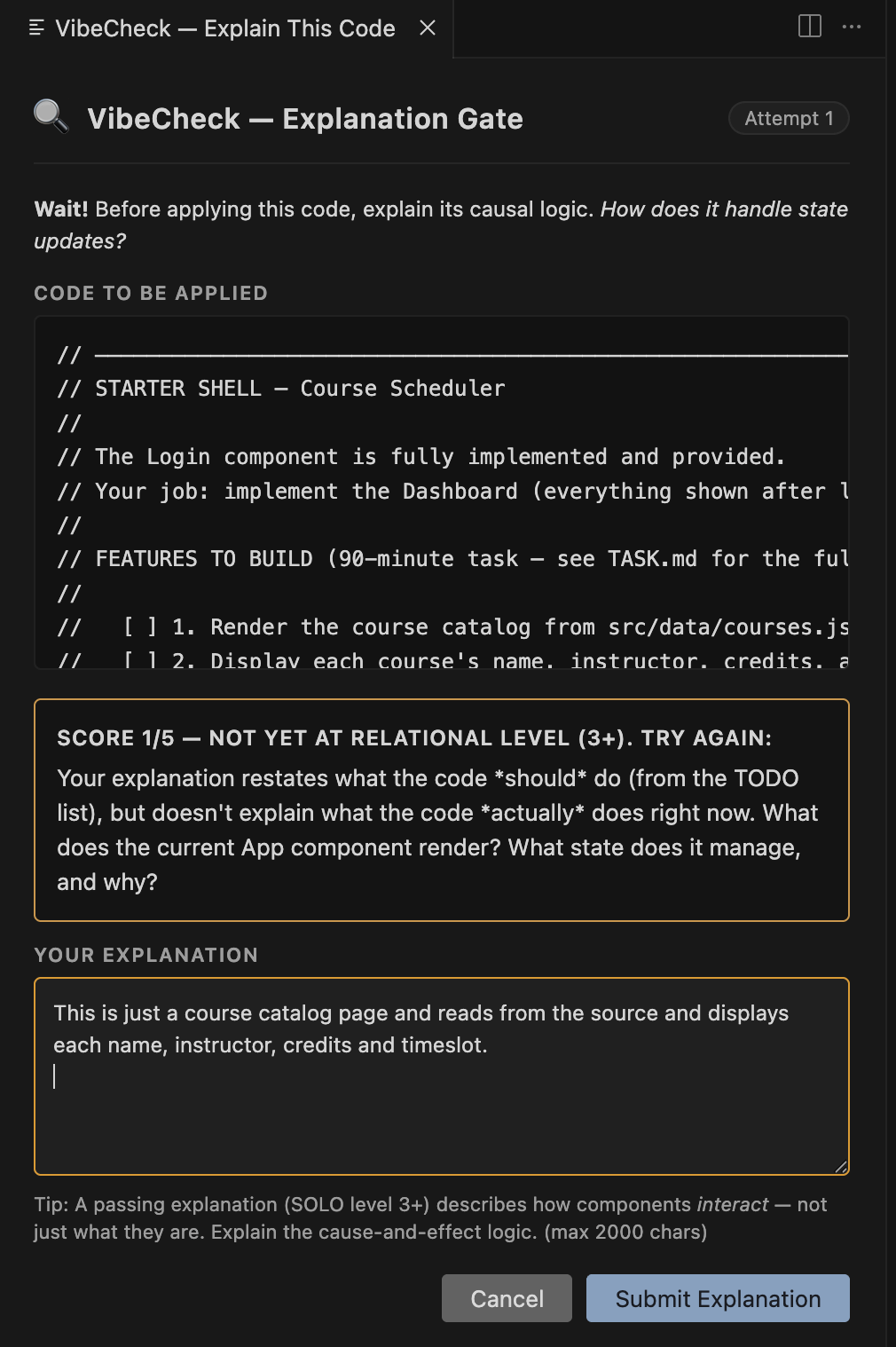}
    \subcaption{SOLO Level~1 (Pre-structural): the Judge identifies that the explanation describes only visible UI behaviour without addressing the underlying mechanism.}
    \label{fig:score1}
  \end{minipage}\hfill
  \begin{minipage}[t]{0.48\linewidth}
    \includegraphics[width=\linewidth]{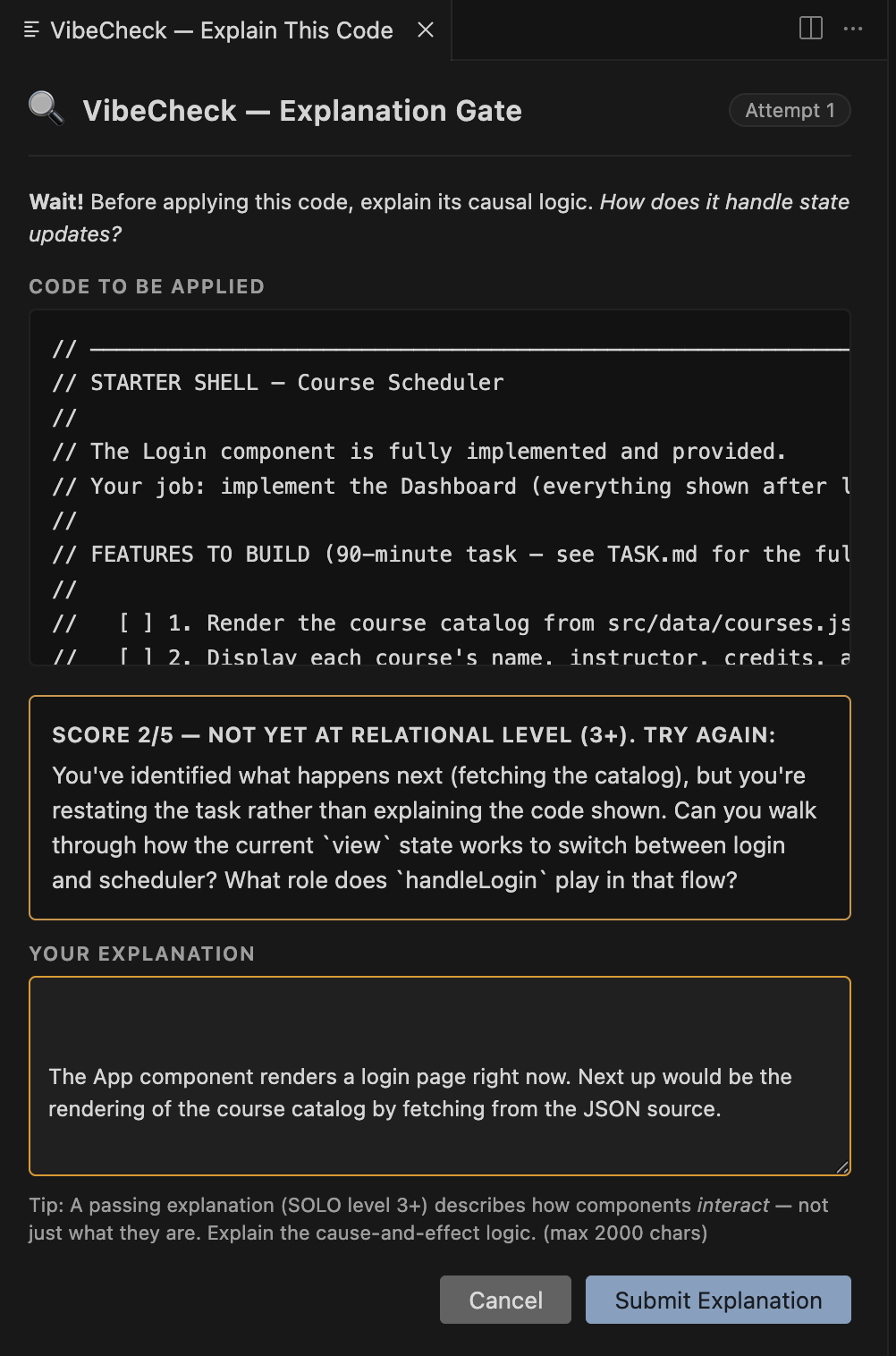}
    \subcaption{SOLO Level~2 (Unistructural): the Judge acknowledges a single correct component but flags the missing causal link between the fetch call and the state update.}
    \label{fig:score2}
  \end{minipage}
  \caption{Example Judge API feedback for sub-threshold explanations. The modal remains locked until a SOLO Level~3 (Relational) response is submitted.}
  \label{fig:judge_feedback}
\end{figure}

The student saw the modal prompt: \textit{``Wait! Before applying this code, explain its causal logic. How does it handle state updates?''} If the Judge returned a score below 3, the modal remained locked, displaying the Judge's Socratic feedback (Figure~\ref{fig:judge_feedback}). This cycle repeated until a passing score was achieved.

The use of the SOLO taxonomy for evaluating student explanations was calibrated with the help of an expert learning designer as the human-in-the-loop, using the process described in \citet{choma2025}.

\begin{table}[ht]
  \caption{SOLO Taxonomy Calibration for Automated Judge Evaluation}
  \label{tab:solo_calibration}
  \small
  \begin{tabularx}{\columnwidth}{c|c|X}
    \toprule
    \textbf{Score} & \textbf{SOLO Level} & \textbf{Student Explanation Example} \\
    \midrule
    1 & Pre-structural & ``I added a button so that when you click it the course is added to the list and it updates.'' \\
    \hline
    2 & Unistructural & ``It uses the \texttt{enrollCourse} function to send the ID to the database and then changes the UI.'' \\
    \hline
    3 & Relational & ``The \texttt{await} keyword ensures the app waits for the DB confirmation before updating the local \texttt{courses} state, preventing UI desync.'' \\
    \hline
    4 & Ext. Abstract (partial) & ``The \texttt{try/catch} around the fetch prevents the UI from showing stale data if the server call fails, but I'm not sure how it handles a timeout.'' \\
    \hline
    5 & Ext. Abstract & ``It implements optimistic updates by updating the state first but uses a \texttt{try/catch} to rollback the UI if the fetch request fails, ensuring eventual consistency.'' \\
    \bottomrule
  \end{tabularx}
\end{table}

\subsection{Experimental Conditions}
Participants were assigned via stratified randomization, stratified by recruitment channel (Prolific vs.\ UserInterviews.com) to ensure balanced representation of both populations across conditions. Within each channel, participants were randomly assigned to one of three conditions, yielding $n=26$ per group.

\subsubsection{Group A: Manual (Control)}
Participants used the standard VS Code environment. They had access to official React documentation but were strictly forbidden from using Generative AI tools.

\textbf{Enforcement:} The \textit{VibeCheck} Rule Enforcement Engine monitored clipboard events; paste events exceeding 20 characters from external sources triggered a warning flag. Additionally, screen recordings of a random 10\% sample were manually reviewed for compliance.

\subsubsection{Group B: Unrestricted AI (Outsourcing)}
Participants used Cursor IDE with free access to Claude 3.5 Sonnet in the sidebar. The Apply button was enabled instantly. This feature automatically diffs and merges generated code into the active file, mimicking the default vibe coding experience.

\subsubsection{Group C: Scaffolded AI (Offloading)}
Participants used Cursor IDE with the extension active. The Apply button was disabled by default. To unlock it, the Explanation Gate Manager triggered a modal asking: \textit{``Explain the causal logic of this block. How does it handle state updates?''}

\textbf{The Judge:} The student's response was sent to a background GPT-4o agent. We purposefully selected a model family different from the generator (Claude) to minimize self-preference bias in evaluation, as explained in the previous section.

\textbf{Feedback Loop:} If rejected, the student received the feedback (e.g., ``You identified \textit{what} variables are used, but not \textit{why} the effect hook is needed here'') and had to retry.

\subsection{Task Protocol}
The experiment consisted of two distinct phases designed to evaluate the trade-off between the immediate production of software and the long-term acquisition of learner competence.

\subsubsection{Phase 1: Functional Utility}
Over a 90-minute period, participants were tasked with building a ``Student Course Scheduler'' React application. This phase was designed to measure functional utility. The application requirements included:
\begin{enumerate}
    \item \textbf{Authentication:} A functional mock login interface.
    \item \textbf{Data Rendering:} Displaying course metadata from a provided JSON source.
    \item \textbf{Stateful Enrollment:} Implementing a selection system to add courses.
    \item \textbf{Conflict Logic:} Algorithmic detection to prevent overlapping schedules.
\end{enumerate}

During this phase, the Rule Enforcement Engine and Explanation Gate Manager were active. For Group C, every attempt to integrate logic triggered the metacognitive friction loop, requiring a passing explanation before code could be merged. 

\textbf{Primary Metric:} The \textit{Functional Utility Score} (0--100), calculated via a headless Puppeteer/Jest test harness that executed 12 end-to-end assertions. Each assertion was weighted equally (8.33 points).

\begin{table}[ht]
    \small
    \caption{12 Component Assertions of the Functional Utility Score for a Total of 100 Points}
    \label{tab:func_util}
    \begin{tabularx}{\columnwidth}{l|X}
        \toprule
        \textbf{ID} & \textbf{Assertion Description} \\
        \midrule
        A-01 & User can log in using mock credentials and reach the dashboard. \\
        A-02 & Course catalog successfully fetches and renders 20+ items from JSON. \\
        A-03 & Course list displays correct metadata (Instructor, Credits, Time). \\
        A-04 & Clicking ``Enroll'' adds a course to the user's sidebar state. \\
        A-05 & Duplicate enrollment in the same course is blocked by state logic. \\
        A-06 & The system identifies time-overlaps between two distinct courses. \\
        A-07 & A warning modal appears when a conflict is detected. \\
        A-08 & Total credit count updates dynamically as courses are added/removed. \\
        A-09 & Enrollment persistence: data remains after simulated page refresh. \\
        A-10 & Search functionality filters the course catalog by name/instructor. \\
        A-11 & Conflict detection logic correctly handles back-to-back classes. \\
        A-12 & Final schedule export generates a valid summary string. \\
        \bottomrule
    \end{tabularx}
\end{table}

\subsubsection{Phase 2: Corrective Competence}
Immediately following Phase 1, we simulated an abstraction leak to measure the participants' \textit{Corrective Competence}. AI access was revoked for all groups by disabling the Cursor sidebar's AI integration via a configuration flag pushed silently by the experimenter; the \textit{VibeCheck} extension simultaneously disabled its gate overlay. Compliance was verified through real-time screen sharing. We then used a git patch to inject a logic bomb into their functioning codebase.

We introduced a race condition in the \texttt{enrollCourse} function by stripping the \texttt{await} keywords from asynchronous database calls and deleting the optimistic UI rollback logic. Under simulated network latency, the UI would optimistically show a course as added, but the underlying state would fail to sync, causing the course to vanish upon browser refresh (a ``ghost course''). Participants were given the final 30 minutes to identify the cause of the ghost courses and manually implement a fix without AI assistance.

It is important to distinguish between general debugging and the measurement of \textit{Corrective Competence}. To ensure we were measuring the latter, the logic bomb was not a novel bug, but a structural regression of the participant's own existing code. Because all participants in Groups B and C had successfully implemented a functional enrollment system in Phase 1 (verified by the test harness), they had all previously ``owned'' a version of the code where the asynchronous \texttt{await} and rollback logic were present. Thus, the Phase 2 task did not require the discovery of a new architectural pattern, but rather the recognition and restoration of a logic flow they had ostensibly integrated. Failure to resolve this regression served as a direct proxy for the epistemic debt accumulated during the initial construction.

\textbf{Primary Metric:} \textit{Corrective Competence}, measured as a binary Repair Success (resolved vs. unresolved) and Time-to-Fix.

\begin{table}[ht]
  \caption{The Phase 2 ``Logic Bomb'' Regression}
  \label{tab:logic_bomb}
  \small
  \begin{tabularx}{\columnwidth}{X|X}
    \toprule
    \textbf{Original (Functional)} & \textbf{Injected Regression (Bug)} \\
    \midrule
    \texttt{const res = await fetch('/enroll');} & \texttt{const res = fetch('/enroll');} \\
    \texttt{if (res.ok) \{ updateUI(); \}} & \texttt{updateUI(); // Optimistic} \\
    \texttt{else \{ rollback(); \}} & \texttt{// No rollback logic} \\
    \bottomrule
  \end{tabularx}
\end{table}

\section{Results}

Our analysis evaluates the experimental data along two dimensions: immediate software production (Utility) and long-term learner competence (Corrective Competence).

\subsection{RQ1: Functional Utility and Velocity}
To evaluate whether the metacognitive friction of the Explanation Gate significantly impeded immediate software production, we analyzed both the Phase 1 Functional Utility Scores and the total time spent on the construction task.

\subsubsection{Functional Utility}
We verified the assumptions for parametric testing; normality was confirmed using a Shapiro-Wilk test ($W = 0.98, p = .42$), and homogeneity of variance via Levene's test ($F(2, 75) = 1.12, p = .33$). A one-way ANOVA revealed a highly significant effect of condition on utility ($F(2, 75) = 48.2, p < .001, \eta^2 = 0.56$).

As detailed in Table \ref{tab:phase1}, both AI-assisted groups achieved significantly higher utility than the manual control. Post-hoc Tukey HSD tests revealed no significant difference in functional utility between the Unrestricted and Scaffolded groups ($p = .64, d = 0.37$, small effect).

\begin{table}[ht]
  \caption{Phase 1: Functional Utility Scores}
  \label{tab:phase1}
  \begin{tabular}{ccl}
    \toprule
    Condition & Mean Utility Score & SD\\
    \midrule
    Manual (Group A) & 65.2\% & 14.3\\
    Unrestricted AI (Group B) & 92.4\% & 8.1\\
    Scaffolded AI (Group C) & 89.1\% & 9.5\\
    \bottomrule
  \end{tabular}
\end{table}

\subsubsection{Development Velocity and Cognitive Allocation}
To evaluate the velocity cost of metacognitive friction, we compared total time-on-task and the allocation of \textit{Active Thinking} time. We define Active Thinking as the duration spent in manual construction, code comprehension, and, for Group C, the duration of Explanation Gate interventions. 

As shown in Table \ref{tab:time_results}, the progress metric reveals that the Manual group exhausted the 90-minute limit having satisfied only 42\% of the requirements. In contrast, both AI-enabled groups achieved 100\% completion. 

While the Scaffolded group encountered a significant velocity penalty compared to the Unrestricted group ($p < .001, d = 1.52$, large effect), their overall velocity remained superior to the Manual control. The friction introduced by the Explanation Gate (Median = 14.2 minutes) was effectively subsidized by the generational speed of the LLM. 

\newcolumntype{C}{>{\centering\arraybackslash}X}

\begin{table}[ht]
  \caption{Phase 1: Time on Task and Cognitive Allocation}
  \label{tab:time_results}
  \small
  \begin{tabularx}{\columnwidth}{lCCC}
    \toprule
    Condition & Total Time Mean ($SD$) & Active Thinking Mean ($SD$) & Progress (\%)\\
    \midrule
    Manual (A) & 88.4 (2.1) & 88.4 (2.1) & 42\%\\
    Unrestricted (B) & 48.2 (10.4) & 18.5 (4.2) & 100\%\\
    Scaffolded (C) & 64.6 (11.2) & 36.3 (5.8) & 100\%\\
    \bottomrule
  \end{tabularx}
\end{table}

\subsection{RQ2: Corrective Competence and Epistemic Debt}
In Phase 2, we measured the \textit{Corrective Competence} of participants after AI access was revoked. Repair Success rates for the race-condition logic bomb were analyzed using a Chi-square test of independence, yielding significant results ($\chi^2(2) = 13.8, p = .001, V = 0.42$).

\begin{itemize}
    \item \textbf{Manual (Group A):} 18/26 succeeded (69.2\%).
    \item \textbf{Unrestricted AI (Group B):} 6/26 succeeded (23.1\%).
    \item \textbf{Scaffolded AI (Group C):} 16/26 succeeded (61.5\%).
\end{itemize}

\subsection{RQ3: Interactional Stances}
We investigated the 6 participants in Group B who successfully fixed the bug. These successful participants utilized a mean of 4.2 explanatory prompts per session, compared to 0.4 for those who failed ($t(24) = 5.1, p < .001, d = 2.37$, very large effect). Their interaction logs showed two qualitatively distinct interactional stances, detailed in the following subsection. 

\begin{table*}[ht]
  \caption{Representative Individual Results showing the divergence in Phase 2.}
  \Description{Table showing Phase 1 utility scores against Phase 2 repair results and time to fix across different conditions and participant IDs.}
  \label{tab:individual}
  \begin{tabular}{llcllc}
    \toprule
    ID & Condition & Utility (Ph 1) & Repair Result (Ph 2) & Fix Time & Dominant Stance (Ph 1)\\
    \midrule
    P-12 & Manual & 72\% & Success & 11m 45s & Manual Reasoning\\
    P-23 & Unrestricted & 95\% & Fail & Timeout & Contractor (Directive)\\
    P-35 & Unrestricted & 98\% & Success & 21m 10s & Consultant (Explanatory)\\
    P-48 & Scaffolded & 90\% & Success & 13m 30s & Enforced Consultant\\
    P-55 & Scaffolded & 82\% & Fail & Timeout & Tautological (Low SOLO)\\
    \bottomrule
  \end{tabular}
\end{table*}

\subsection{The Consultant vs. The Contractor}
Our analysis of the interaction logs in Group B (Unrestricted) reveals a critical behavioral predictor of corrective competence. We identified two dominant interactional stances:

\begin{description}
    \item[\textbf{The contractor stance:}] These participants treated the LLM as a low-level worker. Prompts were purely imperative (e.g., ``Make the button blue,'' ``Add the filter''). Screen recording review showed these users rarely scrolled through the generated code before clicking ``Apply.'' This stance maximizes immediate velocity but results in 100\% cognitive outsourcing.
    \item[\textbf{The consultant stance:}] Participants who succeeded in Phase 2 treated the LLM as a senior partner. Their prompts were interrogative (e.g., ``Why did you use a useEffect here?'' or ``Can you explain this state change?'') \cite{denny2023conversing}. These users engaged in self-scaffolding, effectively ``paying'' their epistemic debt voluntarily.
\end{description}

The \textit{VibeCheck} plugin does not change how participants prompt the AI, but it structurally prevents contractor-stance prompting from resulting in cognitive outsourcing: even a directive prompt requires a SOLO Level~3 explanation before its output is applied. This suggests that the future of AI-assisted learning is not just about the quality of the AI's code, but about designing interfaces that intercept the consequences of the contractor stance.

\subsection{Qualitative Perceptions of Friction}
Post-session surveys revealed a bimodal sentiment toward the Explanation Gate. While 72\% of Group C participants initially described the gate as ``annoying'' or ``an obstacle,'' 64\% of those who successfully fixed the Phase 2 logic bomb admitted that the gate was the only reason they knew where to look for the bug. 

One participant (P-48) noted: \textit{``I was going to just click 'Apply' and move on, but the modal made me realize I didn't actually know why the state was updating. It forced me to read the code I just asked the AI to write.''} This supports our hypothesis that metacognitive friction converts passive outsourcing into active offloading. Nevertheless, future designs should balance beneficial friction against annoyance to encourage voluntary adoption in practice.

\subsection{Distribution of Explanation Quality in the Scaffolded Condition}
We analyzed a total of 412 individual explanation attempts recorded across 26 participants in Group C. These attempts occurred over 172 discrete gate encounters, averaging 6.6 encounters per participant, a frequency that aligns with the six major architectural milestones of the task. 

On average, participants required 2.4 attempts ($SD=1.1$) to satisfy the Judge API's requirement for a SOLO Level~3 (Relational) explanation. 

\begin{figure}[ht]
  \centering
  \Description{Distribution of the number of attempts per Explanation Gate encounter across 172 gate encounters in Group C.}
  \includegraphics[width=\linewidth]{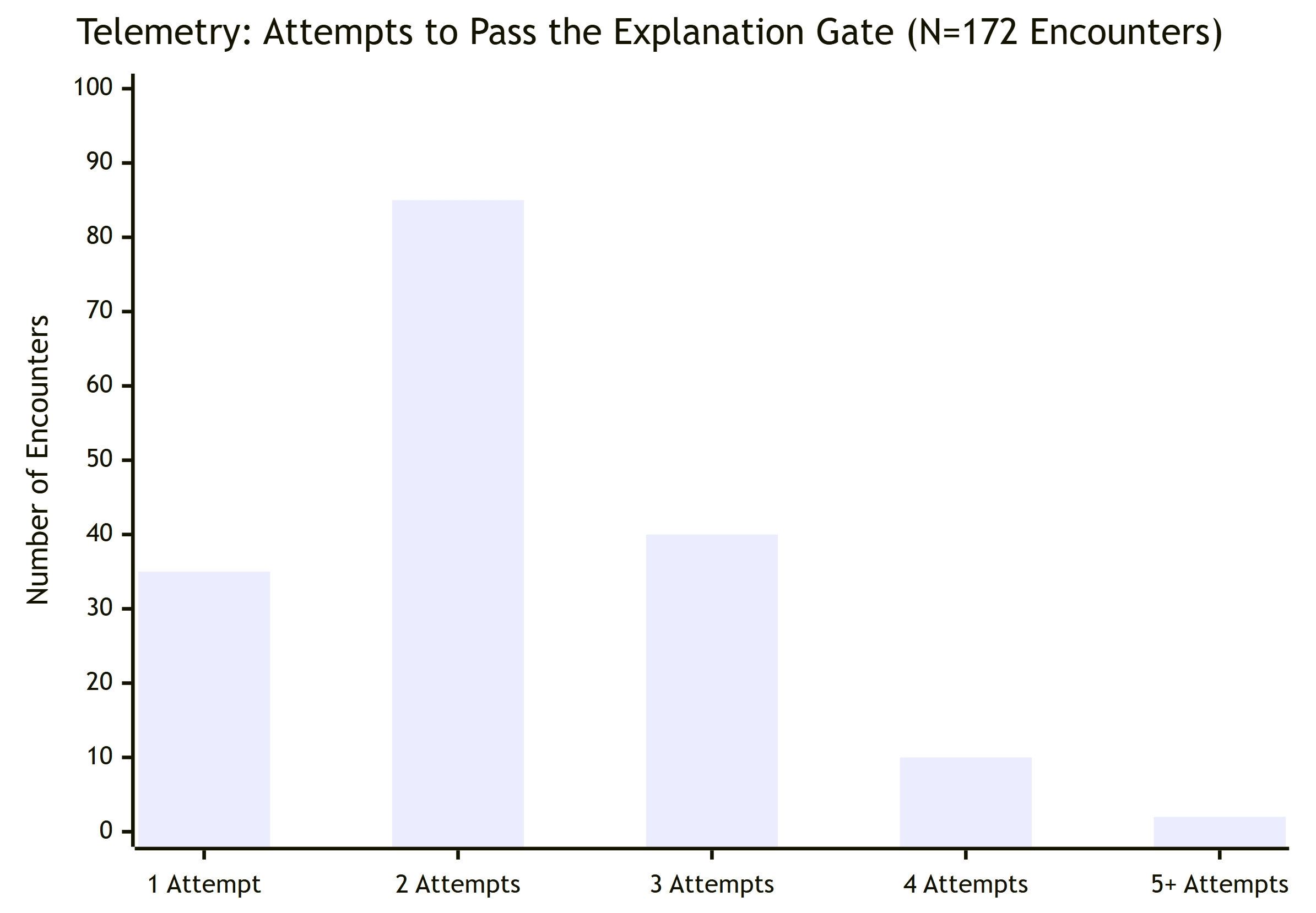}
  \caption{Distribution of the number of attempts per Explanation Gate encounter ($n = 172$ encounters, Group C). The modal value of two attempts indicates that a single round of Socratic feedback was sufficient for most learners to reach SOLO Level~3.}
  \label{fig:attempts_hist}
\end{figure}

\begin{figure}[ht]
  \centering
  \Description{Screenshot of the VibeCheck Explanation Gate at the moment a passing explanation is accepted. The feedback panel reads ``SCORE 5/5 --- Great explanation! Code is being applied...'' The learner's explanation describes how courseMatchesQuery filters courses by name and instructor, how visibleCourses is a memoized filtered result, and how typing in the search input updates the filter state, triggers the memo, and re-renders only the matching rows.}
  \includegraphics[width=0.82\linewidth]{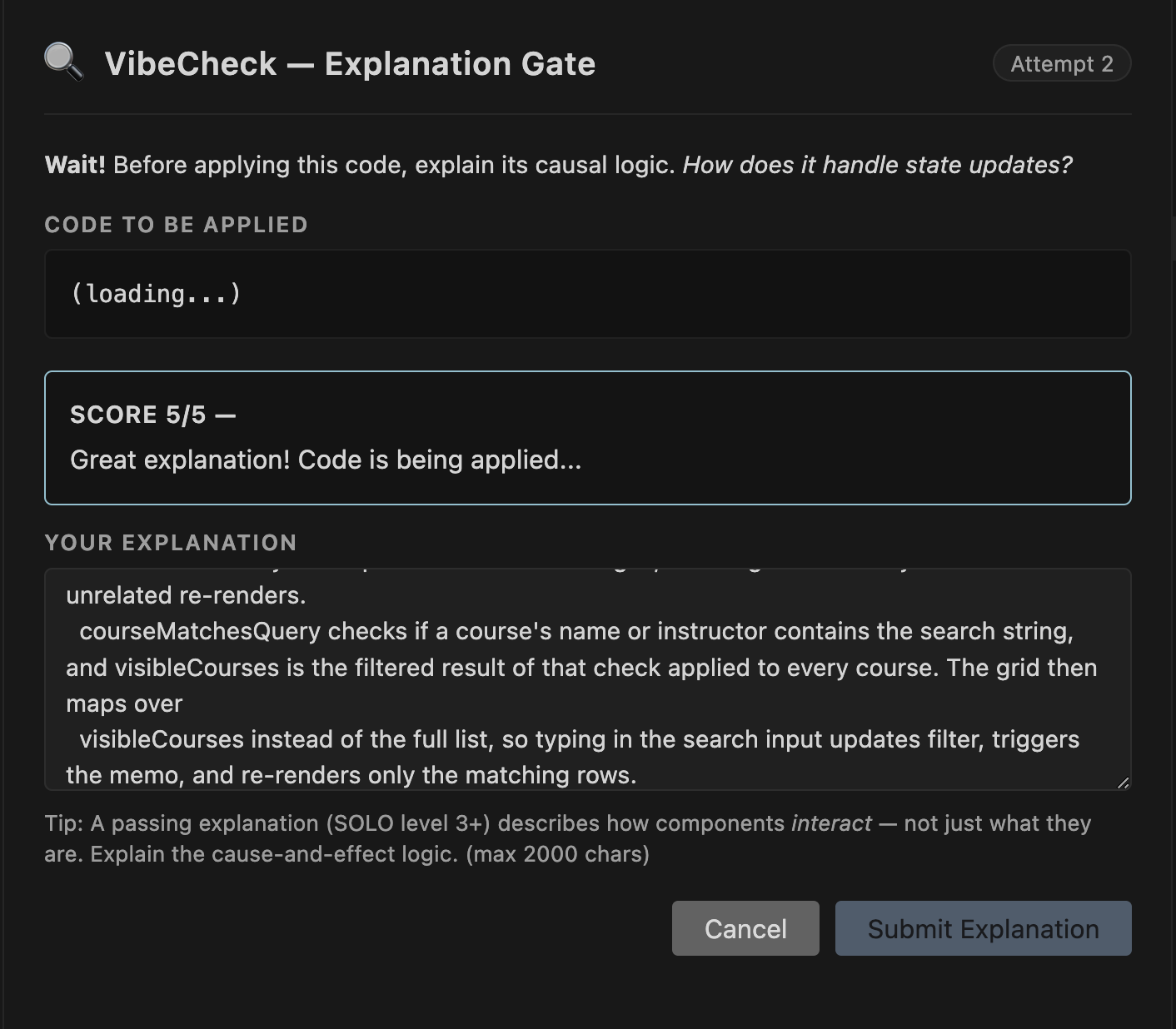}
  \caption{A SOLO Level~3 (Relational) explanation that satisfies the Judge API threshold (Score 5/5). The learner explains the causal chain connecting state, memoization, and re-rendering: the kind of encoding the Gate is designed to produce.}
  \label{fig:gate_pass}
\end{figure}

As shown in Figure \ref{fig:attempts_hist}, the modal peak at two attempts (n=85 encounters) illustrates the impact of the Socratic feedback. The most common failure mode (62\% of rejected attempts) was a tautological pattern, where students repeated the AI's prompt or variable names without explaining causal relationships. The Relational breakthrough typically followed the first round of Judge feedback, shifting the learner from the passive contractor stance to an active consultant stance. Figure~\ref{fig:gate_pass} shows a representative passing explanation (Score 5/5) produced after such a feedback round, illustrating the causal depth the Gate is designed to elicit.

\section{Discussion}

Our findings provide empirical evidence for the existence of epistemic debt in AI-augmented novice programming. By comparing immediate functional utility with subsequent corrective competence, we demonstrate that the efficiency gains of Generative AI can mask a significant collapse in learner comprehension if pedagogical guardrails are not enforced.

The 38-percentage-point gap in repair success between Groups B and C (identical in functional utility during Phase 1) isolates the Explanation Gate as the operative variable. Groups B and C received the same LLM and the same task; the only difference was the presence of metacognitive friction during construction. That this single intervention halved the failure rate in a maintenance task that required no new knowledge suggests the mechanism is cognitive, not procedural: the Gate forced learners to encode the causal logic of AI-generated code rather than consume it passively. Without this encoding, speed gains in Phase 1 were purchased at the cost of the mental representations needed to navigate the same codebase in Phase 2.

\subsection{Mapping the Cognitive Collapse: A CLT Perspective}
To understand the mechanisms behind the observed competence collapse, we mapped our Phase 1 telemetry data to the three components of CLT \cite{sweller1988}.

In the Unrestricted group (B), extraneous load, i.e., the mental effort of managing syntax and IDE state, was effectively reduced to zero. However, so too was germane load: the cognitive effort required to build a durable mental schema. The Explanation Gate in Group C was designed precisely to restore this germane processing. Notably, our qualitative data suggests that Scaffolded participants initially perceived the gate as extraneous friction, indicating that for AI-native learners, pedagogical interventions may be misclassified as system overhead, explaining the system-gaming behaviors observed in specific outliers (e.g., P-55).

We posit that the 38.4\% success rate gap between Groups B and C is primarily attributable to the Explanation Gate forcing germane load, though we cannot fully rule out that the friction itself (rather than the explanation requirement) contributed independently. By requiring a teach-back response passing a SOLO Level~3 threshold, we ensured that learners could not integrate code without first converting the AI's outsourced logic into an offloaded mental representation. Independent neurological support for this mechanism is provided by Kosmyna et al.~\cite{kosmyna2025brain}, whose EEG study of LLM-assisted writers found that ChatGPT users exhibited measurably weaker neural network connectivity than search-engine or unassisted writers, a direct physiological signature of suppressed germane processing that corroborates our behavioral findings. This confirms that metacognitive friction is a necessary catalyst for germane processing in generative environments \cite{chi1989self}.

\subsection{Utility Without Understanding: Debt Accumulation and Code Ownership}
The results from Phase 1 suggest that for novice programmers, unrestricted GenAI access creates a deceptive floor of apparent competence. Group B achieved a functional utility score significantly higher than the Manual group, despite possessing demonstrably less cognitive ownership of the resulting codebase, as Phase 2 confirmed. These participants prioritized the retrieval of terminal outputs over understanding the underlying procedural logic. By vibe coding their way into a degree of complexity they could not sustain, they traded long-term agency for immediate velocity.

We can retrospectively formalize this accumulation of debt as the delta between the immediate \textit{Functional Utility} ($U_f$) enabled by the human-AI dyad and the learner's independent \textit{Corrective Competence} ($C_c$) at time $t$:
\begin{equation}
E_d(t) = U_f(t) - C_c(t)
\end{equation}
Instantiated with our study data: Group B accrued $E_d = 92.4\% - 23.1\% = 69.3$ points of epistemic debt, while Group C, held to account by the Explanation Gate, accrued only $E_d = 89.1\% - 61.5\% = 27.6$ points. Group A, building manually, accrued $E_d = 65.2\% - 69.2\% = -4.0$ points, indicating that genuine construction actually slightly overcomes the initial utility gap through the comprehension it builds. This supports the Yan et al.\ \cite{yan2026beyond} finding that AI support introduces a motivational cost: when the task feels too easy because the AI handles the intrinsic load, the developer stops investing the germane load required to build the mental schemas necessary for $C_c$.

In Phase 2, this apparent competence collapsed. The low repair success rate of the Unrestricted group illustrates what happens when developers treat the AI as a terminal-output machine rather than a collaborative partner: scaling and debugging demand structural understanding that was never internalized. Group B participants treated their own code as an opaque, foreign library.

\subsection{Visual Functionality vs. Structural Comprehension}
The specific nature of the regression injected in Phase 2 (see Table \ref{tab:logic_bomb}) highlights the disconnect between \textit{visual success} and \textit{structural ownership}. By removing the \texttt{await} keyword and the rollback logic, we created a state where the application still appeared to function under ideal conditions but lacked the defensive programming required for robustness.

Participants in the Unrestricted AI group (B) were notably susceptible to this visual illusion of correctness. Because the AI-generated code worked immediately in Phase 1, they never engaged with the underlying asynchronous architecture. When the logic bomb was injected, their troubleshooting logs showed a preoccupation with UI components rather than the data-fetching logic. They could confirm the button existed, but had no mental schema for the promise-based lifecycle of the enrollment event: surface familiarity without structural comprehension. This is the programmatic analog of the ``outside the frontier'' failure mode documented by Dell'Acqua et al.~\cite{dellacqua2026jagged}: in their field experiment, elite consultants who trusted AI on tasks beyond its capability were 19 percentage points less likely to produce correct solutions. Both populations shared the same liability: no internal signal that they had crossed into territory where AI output could not be trusted. 

In contrast, the Scaffolded group (C) was forced to explain this exact lifecycle to the Judge API during construction. By articulating the relationship between the \texttt{fetch} request and the state update, they performed the necessary germane processing to buy back their epistemic debt. When the bug was injected, these participants did not simply debug; they recognized a structural regression in a system they now cognitively owned. This suggests that metacognitive friction acts as an architectural audit, forcing the developer to move beyond the vibe of a working UI to the reality of the underlying implementation.

\subsection{Industry Implications: Realization of Epistemic Debt}
The risks identified in our laboratory setting find a suggestive parallel in recent industry events, lending ecological plausibility to the epistemic debt hypothesis. The Moltbook incident of early 2026 \cite{reuters2026moltbot} serves as a primary case study of the dangers of high utility coupled with low corrective competence. In that event, thousands of developers deployed personal AI agents using generated configurations they did not fully understand. 

While the users successfully orchestrated the functional intent, they outsourced the implementation details, specifically the binding of insecure WebSocket ports. When these systems were scanned by malicious actors, the vibe coders were helpless to intervene. Much like the participants in our Unrestricted group, these developers were effectively senior architects in capability but junior interns in comprehension. Without the foundational mental models required for security and architectural oversight, the functional utility they achieved was fundamentally fragile.

\subsection{The Speed-Comprehension Trade-off}
A common critique of metacognitive interventions is that they impede development speed. However, our results suggest this is a false dichotomy. This critique also rests on a productivity assumption that may itself be flawed: Shen and Tamkin~\cite{shen2026skill} find in a professional programming context that AI-assisted developers showed no statistically significant improvement in task completion time, yet suffered a 17\% reduction in subsequent skill assessment scores ($d = 0.74$). The implication is that unrestricted AI may not even deliver the speed gain it is presumed to trade off against learning. While the Explanation Gate introduced a time penalty of approximately 14 minutes in Group C, this cost was largely subsidized by the generation speed of the LLM.

Crucially, the speed gained through AI assistance in Group C was matched by the comprehension required to maintain it. The \textit{VibeCheck} intervention distinguishes between velocity that is fragile (fast but unmaintainable) and velocity that is sustainable (fast and cognitively owned). By enforcing the teach-back protocol, the tool ensures that the intrinsic load of the task is offloaded rather than outsourced.

\subsection{Self-Scaffolding and Interactional Stances}

The outlier analysis in RQ3 provides a final, critical insight: accumulation of epistemic debt is not solely a function of the tool, but of the user's \textit{interactional stance}. Successful participants in the Unrestricted group naturally adopted a consultant stance. This suggests that while advanced learners may possess the metacognitive discipline to treat AI as a peer, novices require structural friction to prevent them from falling into the contractor stance. Concordant evidence from industry usage data is provided by Anthropic's Economic Index~\cite{anthropic2026economic}, which independently classifies real-world interactions into \textit{automation} (directive, full-delegation) and \textit{augmentation} (feedback loop, task iteration, validation, learning) categories. Their analysis finds that high-tenure users gravitate away from directive automation toward iterative augmentation, precisely the shift from contractor to consultant stance that our intervention enforces structurally for novices. 

The contractor stance observed in Group B corresponds to the emotional dependence on AI interlocutors identified by Yan et al. \cite{yan2026beyond}. These participants did not merely use the tool; they offloaded their critical agency to it. Our Explanation Gate broke this dependence by reintroducing the metacognitive friction necessary to restore the learner's epistemic agency. A complementary taxonomy of stances emerges independently in Shen and Tamkin~\cite{shen2026skill}, who identify six interaction patterns in professional developers learning a new API with AI assistance. Their three low-scoring patterns (AI Delegation, Progressive AI Reliance, and Iterative AI Debugging) correspond directly to the contractor stance, while their high-scoring patterns (Generation-Then-Comprehension, Hybrid Code-Explanation, and Conceptual Inquiry) map onto the consultant. Most tellingly, their Conceptual Inquiry pattern (asking only explanatory questions, never requesting generated code) achieved quiz scores of 65--86\% while matching the completion speed of full delegation, providing direct independent evidence that interactional stance, not tool access, is the determinant of learning outcomes.

Future learning systems must, therefore, move beyond passive code generation toward active metacognitive mediation. By automating the verification of student explanations through an LLM-as-a-Judge framework \cite{choma2025}, we can scale such interventions to meet the demands of learners in the current GenAI era. This imperative extends beyond competence. Lodge and Loble \cite{lodge2026cognitive} argue that education is not merely a cognitive undertaking but a process of formation: novices become thinking practitioners by struggling genuinely with hard problems. Our findings suggest that the Explanation Gate does not only prevent competence collapse; it preserves the conditions under which that formation can occur.

\section{Threats to Validity}

\subsection{Internal Validity}
A primary threat is the Hawthorne effect; participants knew they were in a monitored study, which may have artificially increased their persistence in the Scaffolded group. To mitigate this, we anonymized the Judge's feedback to appear as standard IDE warnings rather than ``test'' feedback. Furthermore, the 90-minute limit for Phase 1 was chosen to simulate the time pressure of a standard technical screen, grounding the velocity results in professional reality. Nevertheless, the limitation remains, despite our mitigation strategies.

\subsection{External Validity}
Our study focused exclusively on novice programmers (students and bootcampers). It is possible that for expert developers, who already possess robust internal mental models, the Explanation Gate would represent purely extraneous load with no germane benefit. The epistemic debt we observed is likely a function of the \textit{novice-AI dyad}, and results may not generalize to experienced programmers who utilize AI for boilerplate tasks rather than core logic. That said, Dell'Acqua et al.~\cite{dellacqua2026jagged} demonstrate that even highly skilled, incentivized knowledge workers (elite BCG consultants) show significant over-reliance failure on outside-frontier tasks, suggesting that the underlying mechanism is not confined to novices. The implication is that scaffolding interventions of the type we study may generalize further up the skill spectrum than our participant sample implies. In either case, a CLT-grounded approach, defining intrinsic, extraneous, and germane loads for each population, can guide the design of appropriate scaffolds. While our study was conducted with a specific novice population, it provides directional support for scaffolding design applicable to experienced programmers as well.

\subsection{Construct Validity}
We used repair success as a proxy for \textit{Corrective Competence}. Critics might argue that this measures debugging skill rather than understanding. However, as noted in Section 3.4.2, because the bug was a regression of the user’s own code integrated moments prior, we believe it remains a valid measure of cognitive ownership.

\subsection{Ecological Validity}
The experiment used a single, controlled task (a React.js Course Scheduler) completed within a two-hour session. Real-world epistemic debt accumulates across months of production work, across diverse technology stacks, and in collaborative codebases. Our study measures an acute form of debt incurred within one session; whether this proxy generalizes to chronic, career-long debt accumulation is an open question. We believe the controlled setting is a strength for causal inference, but future work should replicate the core finding in longitudinal field deployments.

\section{Conclusion}
This study has demonstrated that the democratization of software creation via GenAI carries the latent risk of an industry-wide collapse of competence. However, this breakdown is not inevitable. By deliberately introducing metacognitive friction into the development loop, we can leverage the immense productivity gains of LLMs while ensuring that the humans at the helm remain capable of maintaining the complex systems they build. In our experiment, the Explanation Gate reduced epistemic debt by 41.7 points (from 69.3 to 27.6) at a velocity cost of only 14.2 minutes, a cost largely subsidized by the LLM's generation speed. To prevent the mass production of unmaintainable software, we must ensure that the price of velocity is not the loss of comprehension. The \textit{VibeCheck} plugin and the full experimental protocol are openly available as a replication artifact to support this line of inquiry.

\section{Future Work}

The findings of this study open several avenues for investigating the long-term sustainability of AI-assisted learning environments. We identify four primary areas for future research:

\subsection{Adaptive Scaffolding and Instructional Fading}
The current implementation of the \textit{VibeCheck} Explanation Gate utilizes a static threshold (SOLO Level~3). Future work should explore \textit{adaptive friction}, where the complexity of the required explanation fades as the system detects increasing learner competence. Drawing on the ``expertise reversal effect'' \cite{kalyuga2009expertise}, a system that transitionally moves from high-friction scaffolds to frictionless utility could prevent the fragile expert syndrome while eventually granting the developer the high-velocity benefits of an unrestricted environment.

\subsection{Longitudinal Learning and Mental Model Decay}
Our experiment measured corrective competence immediately following a construction task. However, the long-term persistence of schemas formed under metacognitive friction remains unknown. Future longitudinal studies should investigate whether buying back epistemic debt through teach-back protocols leads to higher retention of architectural principles over months of professional practice, or if the consultant stance naturally erodes as users face the time-pressures of industrial production.

\subsection{Collaborative Human-AI-Human Systems}
While this study focused on the programmer-AI dyad, modern software engineering is a social activity. Research is needed to determine how epistemic debt impacts code reviews and collaborative debugging. If a senior developer reviews code authored by a novice vibe coder, does the lack of cognitive ownership in the junior dev create a bottleneck in the team's collective corrective competence? Extending the Explanation Gate to include collaborative teach-forward protocols between humans could mitigate these social bottlenecks.

\subsection{Equity, Diversity, and the AI Experience Gap}
A critical area for future inquiry is the impact of Generative AI on underrepresented groups in Computer Science. Currently, Black and Hispanic students combined earn close to 20\% of CS degrees in the United States \cite{siek2025taulbee,nces2023digest}, yet they often report higher rates of imposter syndrome and lower access to early-career mentorship. Lodge and Loble \cite{lodge2026cognitive} identify a related \textit{metacognitive equity gap}: the cognitive risks of unstructured AI fall disproportionately on novices who lack the domain knowledge and self-regulation skills needed to use AI beneficially, compounding existing disadvantage. Generative AI has the potential to act as a leveling agent; however, if unrestricted AI usage leads to a disproportionate accumulation of epistemic debt among learners with less existing social capital, it may inadvertently widen the competence gap during technical interviews. Future research must quantify whether scaffolded AI tools like \textit{VibeCheck} provide a more equitable pathway to senior-level architectural understanding for students from diverse socio-economic and racial backgrounds.

\section*{Artifact Availability}
  The \textit{VibeCheck} VS Code/Cursor extension, the Course Scheduler experimental task suite (starter scaffold, reference implementation, 12-assertion Puppeteer grading harness, and logic bomb version), and the complete three-condition study replication protocol are openly available at
  \href{https://github.com/sreecharansankaranarayanan/vibecheck}{https://github.com/\allowbreak{}sreecharansankaranarayanan/\allowbreak{}vibecheck}

\bibliographystyle{ACM-Reference-Format}
\bibliography{references}

\end{document}